\documentclass{article}
 
\usepackage{times}
\usepackage{amssymb,amsmath,amsthm} 
\usepackage{mathrsfs} 
 
\newcommand{\ie}{\emph{i.e.}~}
\newcommand{\eg}{\emph{e.g.}~}

\def\B{\mathscr B}

\def\E{\mathcal E} 
\def\F{\mathscr F} 
 
\def\H{\mathcal H}

\def\rr{\mathbb R}
\def\S{\mathbb S} 

\def\U{\mathscr U}

\def\12{{\textstyle\frac12}}
\def\<{\left\langle} 
\def\>{\right\rangle} 
\def\({\left(} 
\def\){\right)}
\def\[{\left[} 
\def\]{\right]} 
\def\lone{\mathsf{L}^{\:\!\!1}}

\def\ltwo{\mathsf{L}^{\:\!\!2}}
\def\linf{\mathsf{L}^{\:\!\!\infty}}
\def\one{\mathop{1\mskip-4mu{\rm l}}\nolimits}
\def\e{\mathop{\mathrm{e}}\nolimits}
\def\d{\mathrm{d}}
\def\i{\mathrm{i}}

\def\slim{\mathop{\hbox{\rm s-}\lim}\nolimits}
\def\wlim{\mathop{\hbox{\rm w-}\lim}\nolimits}
\def\cD{{\cal D}}
\def\cinf{C^{\infty}}
 
\newcommand{\beq}{\begin{equation}}
\newcommand{\eeq}{\end{equation}}
\newtheorem{theoreme}{Theorem}[section] 
\newtheorem{remark}[theoreme]{Remark} 
\newtheorem{lemma}[theoreme]{Lemma} 
\newtheorem{Assumption}[theoreme]{Assumption} 
 
\newtheorem{proposition}[theoreme]{Proposition} 
\newtheorem{definition}[theoreme]{Definition}

\begin{document} 

    
\title{{\Large\textbf{Generalized definition of time delay in scattering theory}}}
  
\author{Christian G\'erard and Rafael Tiedra de Aldecoa}
\date{\small
\begin{quote}
\emph{
\begin{itemize}
\item[] D\'epartement de math\'ematiques, Universit\'e de Paris XI,\\
91\,405 Orsay Cedex France 
\item[] \emph{E-mails:} christian.gerard@math.u-psud.fr and rafael.tiedra@math.u-psud.fr 
\end{itemize}
}
\end{quote}
}
\maketitle

   
\begin{abstract}
We advocate for the systematic use of a symmetrized definition of time delay in scattering
theory. In two-body scattering processes, we show that the symmetrized time delay exists for
arbitrary dilated spatial regions symmetric with respect to the origin. It is equal to the usual
time delay plus a new contribution, which vanishes in the case of spherical spatial regions. We
also prove that the symmetrized time delay is invariant under an appropriate mapping of time
reversal. These results are also discussed in the context of classical scattering theory.
\end{abstract} 
   
\section{Introduction}\label{Intro}

This paper is devoted to the definition of time delay (in terms of sojourn times) in scattering
theory. Its purpose is to advocate for the systematic use of a symmetrized definition of time
delay. Our main arguments supporting this point of view are the following:
\begin{enumerate}
\item[(A)] Symmetrized time delay generalizes to multichannel-type scattering processes. Usual
time delay does not.
\item[(B)] In two-body scattering processes, symmetrized time delay and usual time delay are
equal.
\item[(C)] In two-body scattering processes, symmetrized time delay does exist for arbitrary
dilated spatial regions symmetric with respect to the origin (usual time delay does exist
only for spherical spatial regions \cite{Sassoli/Martin}). It is equal to the usual time delay
plus a new contribution, which vanishes in the case of spherical spatial regions. 
\item[(D)] Symmetrized time delay is invariant under an appropriate mapping of time reversal.
Usual time delay is not.
\end{enumerate}
Our purpose in this paper is to give the precise meaning and the proof of these statements.

Let us first recall the usual and the symmetrized definition of time delay for a two-body
scattering process in $\rr^d$, $d\ge1$. Consider a bounded open set $\Sigma$ in $\rr^d$
containing the origin and the dilated spatial regions $\Sigma_r:=\{rx\mid x\in\Sigma\}$,
$r>0$. Let $H_0:=-\12\Delta$ be the kinetic energy operator in $\H:=\ltwo(\rr^d)$ and let
$H$ be a selfadjoint perturbation of $H_0$ such that the wave operators
$W^\pm:=\textrm{s-}\lim_{t\to\pm\infty}\e^{\i tH}\e^{-\i tH_0}$ exist and are complete (so
that the scattering operator $S:=(W^+)^*W^-$ is unitary). Then one defines for some states
$\varphi\in\H$ and $r>0$ two sojourn times, namely:
\begin{equation*}
T^0_r(\varphi):=\int_{-\infty}^\infty\d t\int_{x\in\Sigma_r}\d^dx
\left|(\e^{-\i tH_0}\varphi)(x)\right|^2
\end{equation*}
and
\begin{equation*}
T_r(\varphi):=\int_{-\infty}^\infty\d t\int_{x\in\Sigma_r}\d^dx
\left|(\e^{-\i tH}W^-\varphi)(x)\right|^2.
\end{equation*}
If the state $\varphi$ is normalized the first number is interpreted as the time spent by the
freely evolving state $\e^{-\i tH_0}\varphi$ inside the set $\Sigma_r$, whereas the second
one is interpreted as the time spent by the associated scattering state $\e^{-\i tH}W^-\varphi$
within the same region. The (usual) time delay of the scattering process with incoming state
$\varphi$ for $\Sigma_r$ is defined as
\beq\label{usual}
\tau_r^{\rm in}(\varphi):=T_r(\varphi)-T^0_r(\varphi).
\eeq
For a suitable initial state $\varphi$, a sufficiently short-ranged interaction and $\Sigma_r$
spherical, the limit of $\tau_r^{\rm in}(\varphi)$ as $r\to+\infty$ exists and is equal to the
expectation value in the state $\varphi$ of the Eisenbud-Wigner time delay operator
\cite{Amrein/Cibils,ACS}. For multichannel-type scattering processes such as $N$-body scattering
\cite{Smith60,Bolle/Osborn,Martin81}, scattering with dissipative interactions \cite{Martin75},
step potential scattering \cite{Amrein/Jacquet} and scattering in waveguides \cite{Tiedra06}, a
definition such as \eqref{usual} for time delay is inappropriate. In such cases time delay of
the form \eqref{usual} do not admit a limit due to the ``non-conservation" of the kinetic
energy. Therefore one has to modify the definition \eqref{usual} by replacing the free sojourn
time $T^0_r(\varphi)$ with the effective free sojourn time
$\12\[T^0_r(\varphi)+T^0_r(S\varphi)\]$ (see \eg \cite[Sec. V.(a)]{Martin81} or
\cite[Sec. 1]{Tiedra06} for details). In two-body scattering, this modified (symmetrized) time
delay takes the form:
\beq\label{symmetrized}
\tau_r(\varphi):=T_r(\varphi)-\12\[T^0_r(\varphi)+T^0_r(S\varphi)\].
\eeq
We stress that this effective time delay generalises to multichannel-type scattering
proces\-ses (\ie its multichannel counterpart admits a limit as $r\to+\infty$), which is not
the case for the time delay \eqref{usual}.

In Section \ref{quantum-reversal} we prove that the time delay $\tau_r(\varphi)$ is invariant
under an appropriate mapping of time reversal which interchanges past and future scattering
data and reverses the direction of time. In Section \ref{time delay}, Theorem
\ref{abstract time delay}, we give a general existence criterion for the limit
$\tau_\Sigma(\varphi):=\lim_{r\to+\infty}\tau_r(\varphi)$. For arbitrary dilated spatial
regions symmetric with respect to the origin, $\tau_\Sigma(\varphi)$ is shown to be equal to
the usual time delay plus a new contribution, which vanishes in the case of spherical spatial
regions (see Remark \ref{whatitmeans}). Transformations properties of $\tau_\Sigma(\varphi)$
under spatial translations are discussed in Remark \ref{translations}. In the case of
scattering by a short-ranged potential we derive stationary formulas for $\tau_\Sigma(\varphi)$
in Section \ref{sec3b}. These results are also discussed in the context of classical scattering
theory in Section \ref{classical-delay}. Section \ref{sec1} contains some technical results on
averaged characteristic functions.

\section{Averaged characteristic functions}\label{sec1}

Let $\Sigma$ be a bounded open set in $\rr^d$ containing $0$. For each $r>0$ we set
$\Sigma_r:=\{rx\mid x\in\Sigma\}$. We shall simply say that $\Sigma$ is {\em star-shaped}
(resp. {\em symmetric}) whenever $\Sigma$ is star-shaped (resp. symmetric) with respect to $0$.
Clearly $\Sigma$ is star-shaped iff $\Sigma_{r_1}\subset\Sigma_{r_2}$ for $0<r_1\leq r_2$.
Moreover to each open star-shaped set $\Sigma$ we can associate a strictly positive continuous
function $\ell_\Sigma$ on $\S^{d-1}$ defined by
\beq\label{starshaped}
\ell_\Sigma(\omega):=\sup\{\mu\ge 0\mid\mu\omega\in\Sigma\}.
\eeq
Conversely to each strictly positive continuous function $\ell$ on $\S^{d-1}$ one can
associate a unique open star-shaped set $\Sigma$ such that $\ell=\ell_\Sigma$.

We shall also consider the following class of spatial regions $\Sigma$ ($\one_\Sigma$ stands
for the characteristic function for $\Sigma$):
\begin{Assumption}\label{I}
$\Sigma$ is a bounded open set in $\rr^d$ containing $0$ and satisfying the
condition
$$
\int_0^{+\infty}\d\mu\[\one_\Sigma(\mu x)-\one_\Sigma(-\mu x)\]=0,\quad\forall x\in\rr^d.
$$
\end{Assumption}
\noindent
If $p\in\rr^d$, then the number $\int_0^{+\infty}\d t\one_\Sigma(tp)$ is the sojourn time
in $\Sigma$ of a free classical particle moving along the trajectory $t\mapsto x(t):=tp$,
$t\ge0$. Clearly if $\Sigma= -\Sigma$ (\ie if $\Sigma$ is symmetric), then $\Sigma$ satisfies
Assumption \ref{I}. Moreover if $\Sigma$ is star-shaped and satisfies Assumption \ref{I}, then
$\Sigma=-\Sigma$.

\begin{lemma}\label{sojourn}
Let $\Sigma$ be a bounded open set in $\rr^d$ containing $0$. Then:
\begin{enumerate}
\item[(a)] The limit
$$
R_\Sigma(x):=\lim_{\varepsilon\searrow0}\Big(\int_\varepsilon^{+\infty}\frac{\d\mu}\mu
\one_\Sigma(\mu x)+\ln\varepsilon\Big)
$$
exists for each $x\in\rr^d\setminus\{0\}$.
\item[(b)] The (even) function $G_\Sigma:\rr^d\setminus\{0\}\to\rr$ given by
$$
G_\Sigma(x):=\12\[R_\Sigma(x)+ R_\Sigma(-x)\]
$$
satisfies
$$
G_\Sigma(x)=G_\Sigma\big({\textstyle\frac x{|x|}}\big)-\ln|x|.
$$
\item[(c)] If $\Sigma$ is star-shaped, then
\beq
G_{\Sigma}(\omega)=\12\[\ln(\ell_\Sigma(\omega))+ \ln(\ell_\Sigma(-\omega))\]
\label{JimiHendrix}
\eeq
for each $\omega\in\S^{d-1}$.
\end{enumerate}
\end{lemma}

\begin{proof}
Let $x\in\rr^d\setminus\{0\}$. Then point (a) follows from the equalities
\begin{align*}
\lim_{\varepsilon\searrow0}\Big(\int_{\varepsilon}^{+\infty}\frac{\d\mu}\mu
\one_\Sigma(\mu x)+\ln\varepsilon\Big)
&=\int_1^{+\infty}\frac{\d\mu}\mu\one_\Sigma(\mu x)
+\lim_{\varepsilon\searrow0}\int_{\varepsilon}^1\frac{\d\mu}\mu
\[\one_\Sigma(\mu x)-1\]\\
&=\int_1^{+\infty}\frac{\d\mu}\mu\one_\Sigma(\mu x)
+\int_0^1\frac{\d\mu}\mu\[\one_\Sigma(\mu x)-1\].
\end{align*}
Furthermore we have for $\lambda>0$
\begin{align*}
R_\Sigma(\lambda x)&=\lim_{\varepsilon\searrow0}\Big(\int_\varepsilon^{+\infty}
\frac{\d\mu}\mu\one_\Sigma(\mu \lambda x)+\ln \varepsilon\Big)\\
&=\lim_{\varepsilon\searrow0}\Big(\int_{\lambda\varepsilon}^{+\infty}
\frac{\d\mu}\mu\one_\Sigma(\mu x)+\ln(\lambda\varepsilon)-\ln\lambda\Big)\\
&=R_\Sigma(x)-\ln\lambda,
\end{align*}
which proves point (b). Finally point (c) follows from a direct computation.
\end{proof}

We give now some properties of the functions $R_\Sigma$ and $G_\Sigma$, which follow easily
from Lemma \ref{sojourn}.

\begin{remark}\label{rem1}
\begin{enumerate}
\item[(a)] Let us consider $\rr_+^*:=]0,+\infty[$ endowed with the multiplication as a Lie group with
Haar measure $\frac{\d\mu}\mu$. Then $R_\Sigma$ is the (renormalized) average of $\one_\Sigma$
with respect to the action of $\rr_+^*$ on $\rr^d$.
\item[(b)] If $\Sigma$ is equal to the unit open ball $\mathcal B:=\{x\in\rr^d\mid|x|<1\}$, then we have
\beq
G_{\mathcal B}(x)=-\ln|x|,\quad x\in\rr^d\setminus\{0\}.
\eeq
\item[(c)] To each set $\Sigma$ one can associate a unique symmetric star-shaped set
$\widetilde\Sigma$ such that 
$$
G_\Sigma= G_{\widetilde\Sigma}.
$$
Indeed it suffices to take the symmetric star-shaped set $\widetilde\Sigma$ defined by the
even, strictly positive, continuous function $\ell$ on $\S^{d-1}$ given by
$\ell(\omega):=\exp(G_\Sigma(\omega))$.
\end{enumerate}
\end{remark}

\section{Symmetrized time delay in classical scattering}\label{classical-delay}

It is known \cite[Sec. 2]{Sassoli/Martin} that time delay in classical scattering (defined in
terms of sojourn times) does exist only for sequences of dilated balls. In the sequel we recall
the definition of the {\em symmetrized time delay} in classical scattering and show its
existence for (more general) sequences of symmetric spatial regions (we are convinced that
sequences $\Sigma_r$ with initial set $\Sigma$ satisfying Assumption \ref{I} is the optimal
case, but we prefer not to treat this case for the sake of simplicity). We also show that the
symmetrized time delay and the usual time delay are equal for sequences of dilated balls.

We adapt our approach from \cite[Sec. 2]{Sassoli/Martin}. In particular we suppose in the
rest of the section that $\Sigma_r:=\{rx\mid x\in\Sigma\}$, where $\Sigma$ is a convex bounded
open set (with smooth boundary) in $\rr^d$ containing $0$. Let $V$ be a real $C^2$-potential
with compact support. A {\em scattering trajectory} for $V$ is a map
$\Phi:\rr\ni t\mapsto\(x(t),p(t)\)\in\rr^d$ solution of $\dot x(t)= p(t)$,
$\dot p(t)=-\nabla V(x(t))$ satisfying $|x(t)|\to\infty$ as $t\to\pm\infty$ and
$E:=\12p^2(t)+V(x(t))>0$. A scattering trajectory has two asymptotic momenta
$p_\pm:=\lim_{t\to\pm\infty}p(t)$, where $|p_\pm|=(2E)^{1/2}=:p$ due to energy
conservation. Let $\widetilde x(t):=x(t)-tp(t)$ and denote by $-t_-$ and $t_+$
($t_\pm>0$) the times at which the particle enters and leaves the region $\Sigma_r$. If $r$
is large enough, then $\Sigma_r$ contains the support of $v$. Therefore $p(\pm t_\pm)=p_\pm$
and
$$
t_\pm=\mp\frac{p_\pm}{p^2}\cdot\(\widetilde x_\pm-x_\pm\),
$$
where $\widetilde x_\pm:=\widetilde x(\pm t_\pm)$ and $x_\pm:=x(\pm t_\pm)$. One can define
three distinct sojourn times. The sojourn time in $\Sigma_r$ of the {\em scattered particle} is
given by
$$
T_r:=t_-+t_+=\frac1{p^2}\(p_-\cdot\widetilde x_--p_+\cdot\widetilde x_+\)
-\frac1{p^2}\(p_-\cdot x_--p_+\cdot x_+\).
$$
The sojourn time in $\Sigma_r$ of the {\em incoming free particle}
$$
\{x^0(t),p^0(t)\}:=(x_-+p_-(t+t_-),p_-)
$$
with incoming momentum $p_-$, entering time $-t_-$ and leaving time $t^0_+>0$ is
$$
T^0_r=t_-+t^0_+=\frac1{p^2}\(p_-\cdot x^0_+-p_-\cdot x_-\),
$$
where $x^0_+:=x^0(t^0_+)$. The sojourn time in $\Sigma_r$ of the {\em outcoming free particle}
$$
\big\{x^{0'}(t),p^{0'}(t)\big\}:=(x_++p_+(t-t_+),p_+)
$$
with outcoming momentum $p_+$, leaving time $t_+$ and entering time $-t^{0'}_-<0$ is
$$
T^{0'}_r=t^{0'}_-+t_+=\frac1{p^2}\big(p_+\cdot x_+-p_+\cdot x^{0'}_-\big),
$$
where $x^{0'}_-:=x^{0'}(-t^{0'}_-)$. The (usual) {\em time delay} for the finite region
$\Sigma_r$ is defined as
$$
\tau^{\rm in}_r:=T_r-T^0_r.
$$
It is known that $\tau^{\rm in}_r$ admits a limit as $r\to+\infty$ only if $\Sigma$ is
a ball \cite[Sec. 2]{Sassoli/Martin}. In this case $\tau^{\rm in}_r$ converges to the
classical analogue $\tau^{\rm cl}$ of the Eisenbud-Wigner time delay \cite{Narnhofer}.
On the other hand one can also define the {\em symmetrized time delay}:
\beq\label{classical_tau_r}
\tau_r:=T_r-\12\big(T^0_r+T^{0'}_r\big).
\eeq

\begin{remark}
Let $\Phi:\rr\ni t\mapsto\(x(t),p(t)\)\in\rr^d$ be a scattering trajectory and
$p_\pm\equiv p_\pm(\Phi)$, $t_\pm\equiv t_\pm(\Phi)$, $x_\pm\equiv x_\pm(\Phi)$,
$\widetilde x_\pm\equiv\widetilde x_\pm(\Phi)$ the associated scattering quantities.
Consider the mapping ${\rm f}$ (of {\em full time reversal})
${\rm f}:\Phi\mapsto\Phi^{\rm rev}$, where $\Phi^{\rm rev}:t\mapsto\(x(-t),-p(-t)\)$.
Then
$$
p_\pm\circ{\rm f}=-p_\mp,\quad t_\pm\circ{\rm f}=t_\mp,\quad
x_\pm\circ{\rm f}=x_\mp,\quad\widetilde x_\pm\circ{\rm f}=\widetilde x_\mp.
$$ 
Furthermore, setting
$$  
\tau^{\rm out}_r:=\tau^{\rm in}_r\circ{\rm f},
$$
we see that
$$
\tau_r=\12\(\tau^{\rm in}_r+\tau^{\rm out}_r\).
$$
Thus $\tau_r$ is the mean value of the usual time delay $\tau_r^{\rm in}$ and
of the time delay $\tau_r^{\rm out}$ corresponding to the time reversed scattering
process. In particular $\tau_r$ is invariant under full time reversal, namely one has
$\tau_r\circ{\rm f}=\tau_r$ since ${\rm f}$ is an involution.
\end{remark}

The symmetrized time delay \eqref{classical_tau_r} can be rewritten as
$$
\tau_r:=\tau_r^{(1)}+\tau_r^{(2)},
$$
where
\begin{align*}
\tau_r^{(1)}&:=\frac1{p^2}\(p_-\cdot\widetilde x_--p_+\cdot\widetilde x_+\),\\
\tau_r^{(2)}&:=\frac1{2p^2}\big(p_+\cdot x_+-p_-\cdot x_--p_-\cdot x^0_++p_+\cdot x^{0'}_-\big).
\end{align*}
Note that only free trajectories enter in the definition of
$\tau_{r}^{(2)}$.
As $r\to+\infty$ (\ie as $t_\pm\to+\infty$) $\tau_r^{(1)}$ converges to $\tau^{\rm cl}$
\cite[Sec. 2]{Sassoli/Martin}. 

Let us now consider the convergence of $\tau_r^{(2)}$ as $r\to+\infty$. Let $y(t):=y_0+ tp_0$
be an arbitrary free trajectory with $p_0\ne0$. Let $y_\mp$ be the entrance and exit points of
$y(t)$ in $\Sigma_r$. Since these points are independent of the parametrization we can assume
that $y(t)= y_0+ t\omega$, $\omega\in\S^{d-1}$. Thus
\begin{equation}\label{rminus}
r^{-1}y_\pm(y_0,\omega,r)= y_\pm(r^{-1}y_0,\omega,1),
\end{equation}
and $y_\pm(0,\omega,1)=\pm\omega d(\pm\omega)$, where $d(\theta)$ is the distance from the
origin to the boundary of $\Sigma$ in the direction $\theta\in\S^{d-1}$. Since $\Sigma$ is open
the functions $y_\pm$ are continuous w.r.t. to $y_0$. Using \eqref{rminus} this implies that
\beq\label{estimate}
y_{\pm}(y_{0}, \omega, r)=\pm r\omega d(\pm \omega)+o(r).
\eeq
Applying \eqref{estimate} to the two free trajectories $x^0(t)$ and $x^{0'}(t)$, we get
\begin{align}
x_-&=-rd(-\widehat p_-)\widehat p_-+o(r),\quad x^0_+=rd(\widehat p_-)\widehat p_-+o(r),
\label{estimate-bis}\\
x_+&=rd(\widehat p_+)\widehat p_++o(r),\quad x^{0'}_-=-rd(-\widehat p_+)\widehat p_++o(r),
\label{estimate-ter}
\end{align}
where $\widehat p_\pm:=p_\pm/|p_\pm|$. If $\Sigma$ is a ball, then the remainder term in
\eqref{estimate} is actually of order $O(r^{-1})$, and
$$
x_-+x^0_+=O(r^{-1}),\quad x_++x^{0'}_-=O(r^{-1}).
$$
It follows that
\beq\label{identity}
\lim_{r\to+\infty}\tau_r^{(2)}=0\quad{\rm and}\quad
\lim_{r\to+\infty}\tau_r=\lim_{r\to+\infty}\tau_r^{\rm in}=\tau^{\rm cl}.
\eeq 
Equations \eqref{identity} show the identity of the usual time delay and of the symmetrized
time delay in the case of spherical spatial regions.

For an arbitrary $\Sigma$, we get from \eqref{estimate-bis}-\eqref{estimate-ter}:
$$
\tau_r^{(2)}
=\frac r{2p}\[d(\widehat p_+)-d(-\widehat p_+)-d(\widehat p_-)+d(-\widehat p_-)\]+o(r).
$$
Clearly one has to impose that $d(\widehat u)-d(-\widehat u)=0$ for all $\widehat u\in\S^{d-1}$
in order to ensure the existence of the limit $\lim_{r\to+\infty}\tau_r^{(2)}$ for all possible
scattering events. In consequence the limit of $\tau_r$ as $r\to+\infty$ does exist only if the
set $\Sigma$ is symmetric.

\section{Symmetrized time delay in quantum scattering}\label{sec2}

\subsection{Sojourn times}\label{quantum-sojourn}

In this section we recall some properties of the (quantum) sojourn times associated to the free
Hamiltonian $H_0=-\12\Delta$ and the full Hamitonian $H$ in $\H=\ltwo(\rr^d)$. We first recall
some definitions.

$\Sigma$ is a bounded open set in $\rr^d$ containing $0$, and $\Sigma_r=\{rx\mid x\in\Sigma\}$.
For each $r\in\rr^*$ we define the characteristic function
$$
\chi_r(x):=\one_\Sigma(\textstyle\frac xr),\quad x\in\rr^d.
$$
We write $\one_{H_0}(\:\!\cdot\:\!)$ for the spectral measure of $H_0$ and $Q$ for the (vector)
position operator in $\H$. We set $\<\:\!\cdot\:\!\>:=\sqrt{1+|\cdot|^2}$.

We will always assume that:
\begin{Assumption}\label{H1}
The wave operators $W^\pm$ exist and are complete. The projections $\chi_{r}(Q)$ are locally
$H$-smooth on $]0,+\infty[\setminus\sigma_{\rm pp}(H)$.
\end{Assumption}
For latter use we introduce the following definition:
\begin{definition}
Let $s\ge0$, then
\begin{equation*}
\cD_s:=\left\{\varphi\in\cD(\<Q\>^s)\mid
\one_{H_0}(J)\varphi=\varphi\textrm{ for some compact set }J\textrm{ in }
]0,+\infty[\setminus\sigma_{\rm pp}(H)\right\}.
\end{equation*}
\end{definition}
\noindent
It is clear that $\cD_s$ is dense in $\H$ and that $\cD_{s_1}\subset\cD_{s_2}$ if
$s_1\ge s_2$.

For $r>0$ and an appropriate scattering state $\varphi\in\ltwo(\rr^d)$, we define the
{\em free sojourn time}
$$
T^0_r(\varphi):=\int_{-\infty}^{+\infty}\d t\left\|\chi_r(Q)\e^{-\i tH_0}\varphi\right\|^2
$$
and the {\em full sojourn time}
$$
T_r(\varphi):=\int_{-\infty}^{+\infty}\d t\left\|\chi_r(Q)\e^{-\i tH}W^-\varphi\right\|^2.
$$
Due to Assumption \ref{H1}, one shows easily that these times are finite if
$\varphi\in\cD_0$. The {\em time delay} of the scattering process with incoming state
$\varphi\in\cD_0$ for $\Sigma_r$ is then defined as
$$
\tau_r^{\rm in}(\varphi):=T_r(\varphi)-T^0_r(\varphi).
$$
Since $S\cD_0\subset\cD_0$ one can also define the \emph{symmetrized time delay} of the
scattering process with incoming state $\varphi\in\cD_0$:
$$
\tau_r(\varphi):=T_r(\varphi)-\12\[T^0_r(\varphi)+T^0_r(S\varphi)\].
$$
Finally we define for each $r>0$ the auxiliary sojourn time $\tau_r^{\rm free}(\varphi)$
(see \cite[Sec. 2.1]{Tiedra06})
\begin{align}\label{tfree}
\tau_r^{\rm free}(\varphi):=\12&\int^0_{-\infty}\d t\(\|\chi_r(Q)\e^{-\i tH_0}\varphi\|^2
-\|\chi_{r}(Q)\e^{-\i tH_0}S\varphi\|^2\)\\
&+\12\int_0^{+\infty}\d t\(\|\chi_{r}(Q)\e^{-\i tH_0}S\varphi\|^2-
\|\chi_{r}(Q)\e^{-\i tH_0}\varphi\|^2\),\nonumber
\end{align}
which is also finite if $\varphi\in\cD_0$.

\begin{lemma}\label{lemmatfree}
Suppose that Assumption \ref{H1} holds and let $\varphi\in\cD_0$ be such that
\beq\label{H-}
\left\|(W^--\one)\e^{-\i tH_0}\varphi\right\|\in\lone(\rr_-,\d t)
\eeq
and
\beq\label{H+}
\left\|(W^+-\one)\e^{-\i tH_0}S\varphi\right\|\in\lone(\rr_+,\d t).
\eeq
Then
$$
\lim_{r\to+\infty}\[\tau_r(\varphi)- \tau^{\rm free}_r(\varphi)\]=0.
$$
\end{lemma}

\begin{proof}
For $t\in\rr$, set
$$
f_-(t):=\left\|\e^{-\i tH}W^-\varphi-\e^{-\i tH_0}\varphi\right\|\quad\textrm{and}\quad
f_+(t):=\left\|\e^{-\i tH}W^+\varphi-\e^{-\i tH_0}S\varphi\right\|.
$$
We know from Hypotheses \eqref{H-} and \eqref{H+} that $f_\pm\in\lone(\rr_\pm)$. Using the
inequality
\beq\label{norm}
\left|\|u\|^2-\|v\|^2\right|\le\|u-v\|(\|u\|+\|v\|),\quad u,v\in\H,
\eeq
we obtain the estimates
\begin{align*}
\left|\|\chi_r(Q)\e^{-\i tH}W^-\varphi\|^2-\|\chi_r(Q)\e^{-\i tH_0}\varphi\|^2\right|
&\le2f_-(t)\|\varphi\|,\\
\left|\|\chi_r(Q)\e^{-\i tH}W^-\varphi\|^2-\|\chi_r(Q)\e^{-\i tH_0}S\varphi\|^2\right|
&\le2f_+(t)\|\varphi\|.
\end{align*}
Since $\slim_{r\to+\infty}\chi_r(Q)=\one$, then the scalars on the l.h.s. above converge to $0$
as $r\to+\infty$. Thus the claim follows from \eqref{tfree} and Lebesgue's dominated
convergence theorem.
\end{proof}

\subsection{Time reversal}\label{quantum-reversal}

We now collect some elementary remarks related to time reversal for the (complete) scattering
system $\{H_0,H\}$.

Time reversal is implemented by the antiunitary involution
$$
\H\ni\varphi\mapsto\overline\varphi.
$$
The Hamiltonian $H$ is invariant under time reversal if
$$
H\overline\varphi=\overline{H\varphi},\quad\varphi\in\cD(H).
$$
In such a case one has the identities $\e^{-\i tH}\overline\varphi=\overline{\e^{\i tH}\varphi}$,
$W^{\pm}\overline{\varphi}=\overline{W^{\mp}\varphi}$ and
\beq\label{eq2}
S\overline\varphi=\overline{S^{-1}\varphi}
\eeq
for each $\varphi\in\H$. Consider the bijection
$$
{\rm f}:\H\to\H,\quad\varphi\mapsto\overline{S\varphi},
$$
which we call {\em full time reversal}. The map ${\rm f}$ corresponds to time reversal for the
{\em full scattering process}, \ie it interchanges past and future scattering data and reverses
the direction of time. Furthermore one sees easily from \eqref{eq2} that ${\rm f}$ is an
antiunitary involution.

In order to give a rigourous interpretation of full time reversal we introduce the space
${\cal E}$ of {\em scattering trajectories}, \ie the space of continuous maps
$$
\rr\ni t\mapsto\Phi(t)\in\H,
$$
such that
$$
\i(\partial_t\Phi)(t)=H\Phi(t)~~\forall t\in\rr~~\textrm{(in the weak sense)}\quad
\textrm{and}\quad\wlim_{t\to\pm\infty}\Phi(t)=0.
$$
The space ${\cal E}$ is invariant under the involution
$$
R:{\cal E}\to{\cal E},\quad(R\Phi)(t):=\overline{\Phi(-t)}.
$$
One can associate to a trajectory $\Phi\in{\cal E}$ a vector $\varphi:=T(\Phi)\in\H$
defined by the constraint
$$
\slim_{t\to-\infty}\(\Phi(t)-\e^{-\i tH_0}\varphi\)=0.
$$
Due to the completeness of the wave operators we know that $T:{\cal E}\to\H$ is bijective,
and we have
\beq\label{pizza}
{\rm f}(\varphi)=\big(TRT^{-1}\big)(\varphi),\quad\varphi\in\H.
\eeq
Equation \eqref{pizza} provides a rigourous meaning to full time reversal as a map
interchanging past and future scattering data and reversing the direction of time.

\begin{lemma}
Assume that $H$ is invariant under time reversal, and set
$$
\tau_r^{\rm out}(\varphi):=(\tau^{\rm in}_r\circ{\rm f})(\varphi).
$$
Then one has the equalities
\beq\label{eq2ter}
\tau_r(\varphi)=\12\[\tau^{\rm in}_r(\varphi)+\tau_r^{\rm out}(\varphi)\]\quad{\rm and}\quad
\tau_r(\varphi)=(\tau_r\circ{\rm f})(\varphi).
\eeq
\end{lemma}

Thus $\tau_r(\varphi)$ is the mean value of the usual time delay $\tau_r^{\rm in}(\varphi)$ and
of the time delay $\tau_r^{\rm out}(\varphi)$ corresponding to the time reversed scattering
process. In particular $\tau_r(\varphi)$ is invariant under full
time reversal. 

\begin{proof}
Since $H_0$ is invariant under time reversal, one gets
$$
T^0_r(\varphi)=T^0_r(\overline\varphi).
$$
This together with time reversal invariance of $H$ yields
$$
T_{r}(\overline{S\varphi})= T_{r}(\varphi).
$$
Thus
$$
\tau^{\rm out}_r(\varphi)=\tau_r^{\rm in}(\overline{S\varphi})
=T_r(\overline{S\varphi})-T^0_r(\overline{S\varphi})=T_r(\varphi)+T^0_r(S\varphi),
$$
which implies the first identity in \eqref{eq2ter}. The second identity follows from the fact
that ${\rm f}$ is an involution.
\end{proof}

\subsection{Time delay}\label{time delay}

In the present section we shall give the proof of the existence of the symmetrized time delay.
We first fix some notation. If $A,B$ are two symmetric operators, then we set for each
$\varphi\in\cD(A)\cap\cD(B)$:
$$
\(\varphi,[A,B]\varphi\):=\(A\varphi,B\varphi\)-\(B\varphi,A\varphi\).
$$
If $q$ is a quadratic form with domain $\cD(q)$, and $S$ is unitary, then we set for each
$\varphi\in\cD(q)\cap S^{-1}\cD(q)$:
$$
\(\varphi,S^*[q,S]\varphi\):=q(S\varphi)-q(\varphi).
$$
If $A$ is an operator with domain $\cD(A)$ and $S$ is unitary, then we define the operator
$S^*[A,S]$ with domain $\cD(A)\cap S^{-1}\cD(A)$ by
$$
S^*[A,S]:=S^*AS-A.
$$
We also recall that the function $G_\Sigma$ was introduced in Section \ref{sec1} and that
$\cD_2\subset\cD(Q^2)\cap\cD(G_\Sigma(P))$. Therefore the quadratic
form $\i[Q^{2}, G_{\Sigma}(P)]$ is well defined on $\cD_{2}$.

The proof of the next proposition can be found in
the appendix.

\begin{proposition}\label{asympt}
Let $\Sigma$ satisfy Assumption \ref{I}. Suppose that Assumption \ref{H1} is verified. Then we
have for all $\varphi\in\cD_2$ the equality
\begin{align}\label{asymptotic}
&\lim_{r\to+\infty}\int_0^{+\infty}\d t\,\big(\varphi,
\big(\e^{\i tP^2/2}\chi_r(Q)\e^{-\i tP^2/2}-\e^{-\i tP^2/2}\chi_r(Q)\e^{\i tP^2/2}\big)
\varphi\big)\nonumber\\
&=-\(\varphi,\i[Q^2,G_\Sigma(P)]\varphi\).
\end{align}
\end{proposition}

We are now in a position to give the proof of our main theorem. It involves the operator
$$
\textstyle A_0:=\12\big(\frac P{P^2}\cdot Q+Q\cdot\frac P{P^2}\big),
$$
which is well-defined and symmetric on $\cD_1$.

\begin{theoreme}\label{abstract time delay}
Let $\Sigma$ satisfy Assumption \ref{I}. Suppose that Assumption \ref{H1} is verified. Let
$\varphi\in\cD_2$ satisfy \eqref{H-}, \eqref{H+} and $S\varphi\in\cD_2$. Then the limit
$\tau_\Sigma(\varphi)=\lim_{r\to+\infty}\tau_r(\varphi)$ exists, and one has
\beq\label{belle_formule}
\textstyle\tau_\Sigma(\varphi)
=-\12\big(\varphi,S^*\big[\i\big[Q^2,G_\Sigma\big(\frac P{|P|}\big)\big],S\big]\varphi\big)
-\(\varphi,S^*[A_0,S]\varphi\).
\eeq
\end{theoreme}

The quadratic form $\i[Q^{2}, G_{\Sigma}(\frac{P}{|P|})]$ and the
operator $A_{0}$ are well defined on $\cD_{2}$, so all
the commutators in the above formula are well defined since $\varphi,S\varphi\in\cD_2$.

\begin{proof}
The expression \eqref{tfree} for $\tau^{\rm free}_r(\varphi)$ can be rewritten as
\begin{align*}
\tau^{\rm free}_r(\varphi)
&=-\12\int_0^{+\infty}\d t\,\big(\varphi,\big(\e^{\i tP^2/2}\chi_r(Q)\e^{-\i tP^2/2}
-\e^{-\i tP^2/2}\chi_r(Q)\e^{\i tP^2/2}\big)\varphi\big)\\
&\quad+\12\int_0^{+\infty}\d t\,\big(S\varphi,\big(\e^{\i tP^2/2}\chi_r(Q)\e^{-\i tP^2/2}
-\e^{-\i tP^2/2}\chi_r(Q)\e^{\i tP^2/2}\big)S\varphi\big).
\end{align*}
Applying Proposition \ref{asympt}, we get
\begin{align*}
\lim_{r\to +\infty}\tau^{\rm free}_r(\varphi)
&=\12\(\varphi,\i[Q^2,G_\Sigma(P)]\varphi\)-\12\(S\varphi,\i[Q^2,G_\Sigma(P)]S\varphi\)\\
&=-\12\(\varphi,S^*\[\i[Q^2,G_\Sigma(P)],S\]\varphi\).
\end{align*}
By Lemma \ref{sojourn}.(b), we have $G_\Sigma(P)=G_\Sigma\big(\frac P{|P|}\big)-\ln|P|$, and
we know from \cite[Sec. 2]{Amrein/Cibils} that
$$
\textstyle\frac\i2[Q^2,-\ln|P|]=A_0,
$$
as quadratic forms on $\cD_2$. This yields
$$
\textstyle\lim_{r\to +\infty}\tau^{\rm free}_r(\varphi)
=-\12\big(\varphi,S^*\big[\i\big[Q^2,G_\Sigma\big(\frac P{|P|}\big)\big],S\big]\varphi\big)
-\(\varphi,S^*[A_0,S]\varphi\).
$$
We conclude by using Lemma \ref{lemmatfree}.
\end{proof}

\begin{remark}\label{whatitmeans}
The second term in Formula \eqref{belle_formule} coincides with the usual value of time
delay; it is equal to the limit (for $\Sigma$ spherical) of $\tau_r^{\rm in}(\varphi)$ as
$r\to+\infty$ (see \cite[Prop. 1]{Amrein/Cibils}). The first term is a new contribution to
time delay determined by the shape of the set $\Sigma$. If $\Sigma$ is spherical, this
contribution vanishes due to Remark \ref{rem1}.(b), and then one gets (under the
hypotheses of Theorem \ref{abstract time delay}) the equality
$$
\lim_{r\to+\infty}\tau_r(\varphi)=\lim_{r\to+\infty}\tau_r^{\rm in}(\varphi).
$$
\end{remark}

\begin{remark}
Under the hypotheses of Theorem \ref{abstract time delay}, the two following facts are true
whenever $\Sigma$ is an open bounded set containing the origin
(see \cite[Sec. 3]{Sassoli/Martin}):
\begin{enumerate}
\item[(a)] The equality $\lim_{r\to+\infty}\[\tau_r(\varphi)-\tau_r^{\rm free}(\varphi)\]=0$
holds.
\item[(b)] The difference
\begin{align*}
&\int_0^{+\infty}\d t\,
\big(S\varphi,\big[\e^{itP^2}\chi_r(Q)\e^{-itP^2}
-\e^{-itP^2}\chi_r(Q)\e^{itP^2},S\big]\varphi\big)\\
&-r\int_0^{+\infty}\d u
\(S\varphi,\[|H_0|^{-1/2}\(\one_\Sigma\big(u{\textstyle\frac P{|P|}}\big)
-\one_\Sigma\big(-u{\textstyle\frac P{|P|}}\big)\),S\]\varphi\)
\end{align*}
remains bounded as $r\to\infty$.
\end{enumerate}
The integrand in the second term in (b) can be written as
$$
\big(S\varphi,|H_0|^{-1/2}\[M(P),S\]\varphi\big),
$$
where
$$
M(x)=|x|\int_0^{+\infty}\d\mu\[\one_\Sigma(\mu x)-\one_\Sigma(-\mu x)\],\quad x\in\rr^d.
$$
The combination of facts (a) and (b) shows that $\tau_r(\varphi)$ can have a limit for
$\varphi$ in a dense set $\E\subset\H$ only if
$$
\big(S\varphi,|H_0|^{-1/2}\[M(P),S\]\varphi\big)=0,\quad\forall\varphi\in\E,
$$
which implies that $\[M(P),S\]=0$. Therefore, if the scattering operator $S$ has no other
symmetry than $[S,P^2]=0$, one has $M(P)= F(P^2)$ for some function $F$, and it follows that
$M\equiv0$ since $M(x)=-M(-x)$. In consequence $\tau_r(\varphi)$ can have a limit for
$\varphi$ in a dense set $\E$ only if $\Sigma$ satisfies Assumption \ref{I}.
\end{remark}

\begin{remark}\label{translations}
One could also consider time delay for sets $\Sigma_r$ translated by a vector $a\in\rr^d$.
Obviously this is equivalent to determining the time delay \eqref{belle_formule} when the
origin of the spatial coordinate system is translated to the point $a$. In this case one has
\begin{align*}
&\varphi\mapsto\varphi_a:=\e^{\i P\cdot a}\varphi,\\
&S\mapsto S_a:=\e^{\i P\cdot a}S\e^{-\i P\cdot a},
\end{align*}
and $\tau_\Sigma(\varphi)$ becomes
$$
\textstyle\tau^a_\Sigma(\varphi)
:=-\12\big(\varphi_a,S_a^*\big[\i\big[Q^2
,G_\Sigma\big(\frac P{|P|}\big)\big],S_a\big]\varphi_a\big)
-\(\varphi_a,S_a^*[A_0,S_a]\varphi_a\).
$$
Using the formulas
\begin{align*}
\textstyle\e^{-\i P\cdot a}\big[Q^2,G_\Sigma\big(\frac P{|P|}\big)\big]\e^{\i P\cdot a}
&=\textstyle\big[Q^2,G_\Sigma\big(\frac P{|P|}\big)\big]
-2a\cdot\big[Q,G_\Sigma\big(\frac P{|P|}\big)\big],\\
\e^{-\i P\cdot a}A_0\e^{\i P\cdot a}&=\textstyle A_0-a\cdot\frac P{P^2},
\end{align*}
one gets
\beq\label{aveca}
\textstyle\tau^a_\Sigma(\varphi)=\tau_\Sigma(\varphi)
+a\cdot\big(\varphi,S^*\big[\i\big[Q,G_\Sigma\big(\frac P{|P|}\big)\big],S\big]\varphi\big)
+a\cdot\(\varphi,S^*[\frac P{P^2},S]\varphi\).
\eeq
Due to its very definition time delay given by Formula \eqref{aveca} is clearly covariant under
spatial translations.
\end{remark}

\subsection{Stationary formulas}\label{sec3b}

In the sequel we suppose that $\Sigma$ satisfy Assumption \ref{I}. We know from Remark
\ref{rem1}.(c) that there exists a symmetric star-shaped set $\widetilde\Sigma$ such that
$$
\tau^\Sigma(\varphi)=\tau^{\widetilde\Sigma}(\varphi),
$$
for $\varphi$ satisfying the hypotheses of Theorem \ref{abstract time delay}. Thus with no loss
of generality we may assume that $\Sigma$ is symmetric and star-shaped. We also assume that the
boundary $\partial\Sigma$ of $\Sigma$ is a $C^2$ hypersurface, so that the functions
$\ell_\Sigma$ and $G_{\Sigma}$ (see Formulas \eqref{starshaped} and \eqref{JimiHendrix})
associated to $\Sigma$ are $C^2$. In such a case one has
$$
\textstyle\frac\i2\big[Q^2,G_\Sigma\big(\frac P{|P|}\big)\big]
=-\12\big[Q\cdot\nabla G_\Sigma\big(\frac P{|P|}\big)
+\nabla G_\Sigma\big(\frac P{|P|}\big)\cdot Q\big]=:B_\Sigma,
$$
as quadratic forms on $\cD_2$ (note that $B_\Sigma$ is a well-defined symmetric operator on
$\cD_1$). Thus we can rewrite $\tau^\Sigma(\varphi)$ as
\beq\label{station}
\tau_\Sigma(\varphi)=-\(\varphi,S^*[B_\Sigma,S]\varphi\)-\(\varphi,S^*[A_0,S]\varphi\).
\eeq
Let $\U:\ltwo(\rr^d)\to\int_{\rr_+}^\oplus\d\lambda\,\ltwo(\S^{d-1})$ be the spectral
transformation for $H_0$, \ie the unitary mapping defined by
$$
(\U\varphi)(\lambda,\omega)=(2\lambda)^{(d-2)/4}(\F\varphi)(\sqrt{2\lambda}\omega),
$$
where $\F$ denotes the Fourier transform. One has
$$
\U H_0\U^{-1}=\int_{\rr_+}^\oplus\d\lambda\,\lambda\quad{\rm and}\quad
\U S\U^{-1}=\int_{\rr_+}^\oplus\d\lambda\,S(\lambda),
$$
where $\{S(\lambda)\}_{\lambda\ge0}\subset\B\big(\ltwo(\S^{d-1})\big)$ is the scattering matrix for
the pair $\{H_0,H\}$. For shortness we shall set
$\varphi(\lambda):=(T\varphi)(\lambda,\:\!\cdot\:\!)\in\ltwo(\S^{d-1})$.

If the interaction $V:=H-H_0$ is a potential sufficiently short-ranged,
then there exists a dense set $\E\subset\H$ such that the hypotheses of Theorem
\ref{abstract time delay} are satisfied for any $\varphi\in\E$ (a precise definition of $V$
and $\E$ can be found in \cite[Prop.2]{Amrein/Cibils}). Furthermore the function
$\lambda\mapsto S(\lambda)$ is strongly continuously differentiable on $\E$, and the second
term in \eqref{station} is equal to the Eisenbud-Wigner time delay for any $\varphi\in\E$:
$$
-\(\varphi,S^*[A_0,S]\varphi\)=-\i\int_0^\infty\d\lambda\(\varphi(\lambda),
S(\lambda)^*\(\textstyle\frac{\d S(\lambda)}{\d\lambda}\)\varphi(\lambda)\)_{\ltwo(\S^{d-1})}
\equiv\(\varphi,\tau_\textsc{e-w}\varphi\).
$$
Let us now consider the first term in \eqref{station}. Since the function
$x\mapsto G_\Sigma\big(\frac x{|x|}\big)$ is homogeneous of degree $0$, one has
$$
\textstyle x\cdot(\nabla G_\Sigma)\big(\frac x{|x|}\big)=0,
$$ 
namely the vector field $(\nabla G_{\Sigma})\big(\frac x{|x|}\big)$ is orthogonal to the
radial direction. In fact a direct calculation shows that
$$
\U B_\Sigma\U^{-1}=\int_{\rr_+}^\oplus\d\lambda\,\lambda^{-1}b_\Sigma(\omega,\partial_\omega),
$$
where $b_\Sigma(\omega,\partial_\omega)$ is a symmetric first order differential operator on
$\S^{d-1}$ with $C^1$ coefficients. Therefore the operator $\U B_\Sigma\U^{-1}$ is
essentially selfadjoint on $\U\cD_1$, and its closure is decomposable in the spectral
representation of $H_0$, \ie
$$
\overline{\U B_\Sigma\U^{-1}}=\U\overline{B_\Sigma}\U^{-1}\equiv
\int^\oplus_{\rr_+}\d\lambda\,B_\Sigma(\lambda).
$$
This yields the equality
\begin{align*}
\tau_\Sigma(\varphi)&=-\int_0^{+\infty}\d\lambda\,\(\varphi(\lambda),
S^*(\lambda)[B_\Sigma(\lambda),S(\lambda)]\varphi(\lambda)\)_{\ltwo(\S^{d-1})}\\
&\qquad\qquad-\i\int_0^{+\infty}\d\lambda\,\(\varphi(\lambda),S(\lambda)^*
\(\textstyle\frac{\d S(\lambda)}{\d\lambda}\)\varphi(\lambda)\)_{\ltwo(\S^{d-1})}.
\end{align*}
In consequence the time delay \eqref{station} is the sum of two contributions, each of these
being the expectation value of an operator decomposable in the spectral representation of
$H_0$.

\section*{Appendix}

\begin{proof}[Proof of Proposition \ref{asympt}]
(i) For any $F\in\linf(\rr^d)$ and $s\in\rr$ one has
\begin{align}
\e^{\i sP^2/2}F(Q)\e^{-\i sP^2/2}&=F(Q+s P),\nonumber\\
\e^{-\i sQ^2/2}F(P)\e^{\i sQ^2/2}&=F(P+s Q),\label{triv2}
\end{align}
which imply the identity
\beq\label{conjugation}
\e^{\i tP^2/2}F(Q)\e^{-\i tP^2/2}=Z_{-1/t}F(tP)Z_{1/t},
\eeq
where $t\in\rr^*$ and $Z_\tau:=\e^{\i\tau Q^2/2}$. Formula (\ref{conjugation}) and the change
of variables $\mu=rt^{-1}$, $\nu=r^{-1}$, lead to the equalities
\begin{align*}
&\int_0^{+\infty}\d t\,\big(\varphi,
\big(\e^{\i tP^2/2}\chi_r(Q)\e^{-\i tP^2/2}-\e^{-\i tP^2/2}\chi_r(Q)\e^{\i tP^2/2}\big)
\varphi\big)\nonumber\\
&=\int_0^{+\infty}\frac{\d\mu}{\nu\mu^2}\(\varphi,\(Z_{-\nu\mu}\chi_\mu(P)Z_{\nu\mu}
-Z_{\nu\mu}\chi_{-\mu}(P)Z_{-\nu\mu}\)\varphi\).
\end{align*}
One has also
$$
\int_0^{+\infty}\frac{\d\mu}{\mu^2}\[\chi_\mu(P)-\chi_{-\mu}(P)\]
=\int_0^{+\infty}\d s\[\chi(sP)-\chi(-sP)\]=0
$$
due to Assumption \ref{I}. Hence the l.h.s of \eqref{asymptotic} can be written as
\beq\label{limit}
K_\infty(\varphi):=\lim_{\nu\searrow0}\int_0^{+\infty}\d\mu\,K_{\nu,\mu}(\varphi),
\eeq
where
\begin{align*}
\textstyle K_{\nu,\mu}(\varphi):=\frac1{\nu\mu^2}&\(\varphi,
\[Z_{-\nu\mu}\chi_\mu(P)Z_{\nu\mu}-\chi_\mu(P)\]\varphi\)\\
&\textstyle-\frac1{\nu\mu^2}
\(\varphi,\[Z_{\nu\mu}\chi_{-\mu}(P)Z_{-\nu\mu}-\chi_{-\mu}(P)\]\varphi\).
\end{align*}
(ii) To prove the statement, we shall show that one may interchange the limit and the integral
in \eqref{limit}, by invoking Lebesgue's dominated convergence theorem. This will be done in
(iii) below. If one assumes that this interchange is justified for the moment, then direct
calculations give
\begin{align}\label{hohoho}
K_\infty(\varphi)
&=\int_0^{+\infty}\d\mu\,\frac\d{\d\nu}\,K_{\nu,\mu}(\varphi)\Big|_{\nu=0}\nonumber\\
&=-\12\int_0^{+\infty}\frac{\d\mu}\mu
\(\varphi,\i\([Q^2,\chi_\mu(P)]+[Q^2,\chi_{-\mu}(P)]\)\varphi\).
\end{align}
Due to Lemma \ref{sojourn}.(a) we have
\begin{align*}
&\int_0^{+\infty}\frac{\d\mu}\mu\(\varphi,[Q^2,\chi_\mu(P)]\varphi\)\\
&=\lim_{\varepsilon\searrow0}\int_\varepsilon^{+\infty}\frac{\d\mu}\mu
\[\(Q^2\varphi,\chi_\mu(P)\varphi\)-\(\chi_\mu(P)\varphi,Q^2\varphi\)\]\\
&=\lim_{\varepsilon\searrow0}\big[
\big(Q^2\varphi,\big(\textstyle{\int_\varepsilon^{+\infty}\frac{\d\mu}\mu}\,\chi_\mu(P)
+\ln\varepsilon\big)\varphi\big)
-\big(\big(\textstyle{\int_\varepsilon^{+\infty}\frac{\d\mu}\mu}\,\chi_\mu(P)
+\ln\varepsilon\big)\varphi,Q^2\varphi\big)\big]\\
&=\(\varphi,[Q^2,R_\Sigma(P)]\varphi\).
\end{align*}
This together with \eqref{hohoho} lead to the desired equality, that is
$$
K_\infty(\varphi)=-\12\(\varphi,\i[Q^2,R_\Sigma(P)+R_\Sigma(-P)]\varphi\)
=-\(\varphi,\i[Q^2,G_\Sigma(P)]\varphi\).
$$

(iii) To apply Lebesgue's dominated convergence theorem to \eqref{limit} we need
to bound $K_{\nu,\mu}(\varphi)$ uniformly in $\nu$ by a function in $\lone(\rr_+,\d\mu)$.
We do this separately for $\mu\leq1$ and for $\mu\ge1$.

We begin with the case $\mu\leq1$. Write $K_{\nu,\mu}(\varphi)$ as
$$
K_{\nu,\mu}(\varphi)=F_{\nu,\mu}(\varphi)-F_{\nu,-\mu}(\varphi),
$$
where
\begin{align*}
F_{\nu,\mu}(\varphi)
&\textstyle=\frac1{\nu\mu^2}\[\(Z_{\nu\mu}\varphi,\chi_\mu(P)Z_{\nu\mu}\varphi\)
-\(\varphi,\chi_\mu(P)\varphi\)\]\\
&\textstyle=\frac1\mu\big(\big(\frac{Z_{\nu\mu}-\one}{\nu\mu}\big)\varphi,
\chi_\mu(P)Z_{\nu\mu}\varphi\big)+\frac1\mu\big(\chi_\mu(P)\varphi,
\big(\frac{Z_{\nu\mu}-\one}{\nu\mu}\big)\varphi\big).
\end{align*}
Due to the spectral theorem, we have
\beq\label{eq5}
\textstyle\big\|\big(\frac{Z_{\pm\nu\mu}-\one}{\nu\mu}\big)\varphi\big\|
\le{\rm Const.}\big\|\<Q\>^2\varphi\big\|.
\eeq
Let $0<\ell<\12$, then $|P|^{-\ell}\<Q\>^{-2}$ is bounded (after conjugation by a Fourier
transform this follows from the fact that $|Q|^{-\ell}$ is $P^2$-bounded
\cite[Prop. 2.28]{Amrein81}). Since $\Sigma$ is bounded, we have
$$
|\mu^{-1}\xi|^{\ell}|\chi_{\pm\mu}(\xi)|\leq{\rm Const.}
$$
Therefore
\begin{align}\label{eq6}
\mu^{-1}\left\|\chi_{\pm\mu}(P)\varphi\right\|
&=\mu^{\ell-1}\big\||\mu^{-1}P|^\ell\chi_{\pm\mu}(P)
|P|^{-\ell}\<Q\>^{-2}\<Q\>^2\varphi\big\|\nonumber\\
&\leq{\rm Const.}\,\mu^{\ell-1}\big\|\<Q\>^2\varphi\big\|,
\end{align}
and
\beq\label{ptit_frere}
\mu^{-1}\left\|\chi_{\pm\mu}(P)Z_{\pm\nu\mu}\varphi\right\|
\leq{\rm Const.}\,\mu^{\ell-1}\big\|\<Q\>^2\varphi\big\|.
\eeq
From \eqref{eq5}, \eqref{eq6} and \eqref{ptit_frere} we get the estimates
$$
|F_{\nu,\pm\mu}(\varphi)|\leq{\rm Const.}\,\mu^{\ell-1}\big\|\<Q\>^2\varphi\big\|^2.
$$
Thus we have
\beq\label{abba}
|K_{\nu,\mu}(\varphi)|\leq{\rm Const.}\,\mu^{\ell-1}\big\|\<Q\>^2\varphi\big\|^2,
\eeq
which shows that $K_{\nu,\mu}(\varphi)$ is bounded uniformly in $\nu$ by a function in
$\lone([0,1],\d\mu)$.

We consider now the case $\mu\ge1$. Since $\varphi=\one_J(H_0)\varphi$ for some compact
set $J$, there exists $\mu_0\ge0$ such that
$$
(\varphi,\chi_\mu(P)\varphi)=(\varphi,\chi_{-\mu}(P)\varphi)
=(\varphi,\varphi),\quad\forall\mu\ge\mu_0.
$$
Hence for $\mu\ge\mu_0$, we have
\begin{align}\label{eq01}
|K_{\nu,\mu}(\varphi)|&
=\textstyle\frac1{\nu\mu^2}\left|\(\varphi,Z_{-\nu\mu}\chi_\mu(P)Z_{\nu\mu}\varphi\)
-\(\varphi,Z_{\nu\mu}\chi_{-\mu}(P)Z_{-\nu\mu}\varphi\)\right|\nonumber\\
&=\textstyle\frac1{\nu\mu^2}\left|\(\|\chi_\mu(P)Z_{\nu\mu}\varphi\|^2-\|Z_{\nu\mu}\varphi\|^2\)
-\(\|\chi_{-\mu}(P)Z_{-\nu\mu}\varphi\|^2-\|Z_{-\nu\mu}\varphi\|^2\)\right|\nonumber\\
&\le\textstyle\frac2{\nu\mu^2}\|\varphi\|\[\|(\chi_\mu(P)-\one)Z_{\nu\mu}\varphi\|
+\|(\chi_{-\mu}(P)-\one)Z_{-\nu\mu}\varphi\|\],
\end{align}
where we have used \eqref{norm} in the last step. To bound the r.h.s. of \eqref{eq01}
we will use the following identity, which is an easy consequence of \eqref{triv2}:
\beq\label{eq8}
\[F(P+sQ)-F(P)\]\varphi
=\12\int_0^s\d\tau\[2(\nabla F)(P+\tau Q)\cdot Q-\i(\Delta F)(P+\tau Q)\]\varphi,
\eeq
where $F$ is any bounded function in $\cinf(\rr^d)$ with bounded derivatives.

Let $F\in\cinf(\rr^d)$ with $F\equiv1$ near infinity, $F\equiv0$ near $0$ be such that 
$$
F(Q)[\chi(Q)-\one]=\chi(Q)-\one.
$$
Then we have
\beq\label{eq9}
\textstyle\|[\chi_{\pm\mu}(P)-\one]Z_{\pm\nu\mu}\varphi\|
\le\big\|F\big(\frac{\pm P}\mu\big)Z_{\pm\nu\mu}\varphi\big\|
=\big\|F\big(\frac{\pm P}\mu+\nu Q\big)\varphi\big\|
\eeq
due to \eqref{triv2}. From \eqref{eq9} and the fact that
$$
\textstyle F\big(\frac{\pm P}\mu\big)\varphi=0,\quad\forall\mu\ge\mu_0,
$$
we get for $\mu\ge\mu_0$
$$
|K_{\nu,\mu}(\varphi)|\le\textstyle\frac2{\nu\mu^2}\|\varphi\|\left\{
\big\|\big[F\big(\frac P\mu+\nu Q\big)-F\big(\frac P\mu\big)\big]\varphi\big\|
+\big\|\big[F\big(\frac{-P}\mu+\nu Q\big)-F\big(\frac{-P}\mu\big)\big]\varphi\big\|
\right\}
$$
Moreover one has
$$
\textstyle\big[F\big(\frac{\pm P}\mu+\nu Q\big)-F\big(\frac{\pm P}\mu\big)\big]\varphi
=\12{\displaystyle\int_0^\nu}\d\tau\big[2(\nabla F)
\big(\frac{\pm P}\mu+\tau Q\big)\cdot Q-\i(\Delta F)\big(\frac{\pm P}\mu+\tau Q\big)\big]\varphi
$$
due to \eqref{eq8}. Therefore we have for $\mu\ge\mu_0$
\beq
|K_{\nu,\mu}(\varphi)|\le{\rm Const.}\,\mu^{-2}\left\|\<Q\>\varphi\right\|^2.
\label{eq10}
\eeq
The combination of \eqref{abba} and \eqref{eq10} shows that $K_{\nu,\mu}(\varphi)$ is bounded
uniformly in $\nu$ by a function in $\lone([1,+\infty[,\d\mu)$.
\end{proof}

\section*{Acknowledgements} 
 
R.T.d.A. thanks W. O. Amrein and P. Jacquet for their helpful remarks, and he also thanks the
Swiss National Science Foundation for financial support.


\end{document}